\documentclass[notitlepage,superscriptaddress,showpacs,a4paper,nobalancelastpage,prb,twocolumn]{revtex4-1}%
\usepackage{amsfonts}
\usepackage{subfigure}
\usepackage{amsmath}
\usepackage{amssymb}
\usepackage{graphicx,epstopdf}
\usepackage{appendix}
\usepackage{array}
\usepackage{color}
\usepackage{my_symbols}
\usepackage[normalem]{ulem}
\usepackage{CJK}
\usepackage{float}
\providecommand{\U}[1]{\protect\rule{.1in}{.1in}}

\newcommand\rmv{\bgroup\markoverwith {\textcolor{red}{\rule[0.5ex]{2pt}{0.4pt}}}\ULon}

\begin{document}

\begin{CJK*}{UTF8}{gbsn} 
\title{Intrinsic magnetoresistance in metal films on ferromagnetic insulators}
\author{Vahram L. Grigoryan}
\fudan
\author{Wei Guo}
\fudan
\author{Gerrit E. W. Bauer}
\affiliation{Kavli Institute of NanoScience, Delft University of Technology, Delft, The Netherlands}
\affiliation{Institute for Materials Research and WPI-AIMR, Tohoku University, Sendai, Japan}
\author{Jiang Xiao (萧江)}
\email[Corresponding author:~]{xiaojiang@fudan.edu.cn}
\fudan
\begin{abstract}
We predict a magnetoresistance induced by the interfacial Rashba spin-orbit coupling in normal metal$\big|$ferromagnetic insulator bilayer. It depends on the angle between current and magnetization directions identically to the ``spin Hall magnetoresistance'' mechanism caused by a combined action of spin Hall and inverse spin Hall effects. Due to the identical phenomenology it is not obvious whether the magnetoresistance reported by Nakayama \etal \cite{nakayama_2013} is a bulk metal or interface effect. The interfacial Rashba induced magnetoresistance may be distinguished from the bulk metal spin Hall magnetoresistance by its dependence on the metal film thickness.
\end{abstract}

\pacs{}
\maketitle
\end{CJK*}
The spin-orbit interaction (SOI) couples the charge and spin degrees of freedom of the ~electron. \cite{fabian_2007}  In particular, the SOI-induced spin Hall effect (SHE), \cite{dyakonov_1971,hirsch_1999} which converts an electric current into a transverse spin current, provides an electrical method of generating pure spin currents. The inverse spin Hall effect (ISHE) converts a spin current into a transverse electric current or voltage, thereby detects pure spin currents. Both SHE and ISHE are bulk material effects. The functionality of spintronic devices is often governed by the interfaces between different materials. \cite{parkin_1990} At interfaces structural inversion symmetry is broken, thereby allowing the emergence of a Rashba-type of spin-orbit interaction. Recently evidences has surfaced that the Rashha term can be substantial and play a critical role in controlling interfacial electronic states and magnetization textures. Interfacial Rashba spin-orbit interactions (IR-SOI) can cause spin Hall-like effects in normal metal thin films without bulk spin-orbit coupling (such as Cu or Al). \cite{wang_2013}

The SOI can generate spin-orbit $\text{ }$ torques \cite{miron_2011,liu_2011a,liu_2012,kureba_2014,manchon_2008,manchon_2009,manchon_2012,kim_2013,
cher_2010} in a single ~ ferromagnetic film (FM) without an external polarizer \cite{manchon_2008,manchon_2009,manchon_2012,kim_2013} and dilute magnetic semiconductors \cite{cher_2010} with different top and bottom layers as well as in ferromagnetic metals. \cite{miron_2011,liu_2011a} However, the microscopic origin of the current-driven magnetization dynamics is still controversial. An alternative interpretation of the experimental observations relies on the SHE \cite{dyakonov_1971,liu_2011a,hirsch_1999,wunderlich_2005} occurring in the NM\big|FM structures. \cite{liu_2011a} In studies on bilayer or multilayer structures \cite{miron_2011,liu_2012} with broken inversion symmetry many other phenomena, \cite{engel_2007,liu_2011a} like magnetization switching or domain wall motion, have been explained by either SHE or interfacial spin-orbit torques. \cite{liu_2011b,haazen_2013} Often the bulk metal \cite{liu_2011a} and Rashba SOI \cite{manchon_2012} depend identically on the angle between applied current and magnetization direction, \cite{manchon_2012} so it is$\text{ }$ difficult to disentangle their relative importance. However, because the SHE is a bulk effect, while the Rashba field is an interface effect, systematic studies of the thickness dependence may help distinguishing their contributions. \cite{kim_2013}

A straightforward interpretation of experiments on magnetization switching and domain wall motion is hindered by the strongly non-linear magnetization dynamics involved. The magnetoresistance provides a much more straightforward access to the effects of SOI on transport. 

The ferrimagnetic insulator yttrium iron garnet (Y$_3$Fe$_5$O$_{12}$, YIG) has recently attracted the interest of the spintronics community. \cite{wu_2013} Since no current flows through the ferromagnet the choice of YIG simplifies the interpretation of transport in bilayers. Nakayama \textit{et al.} \cite{nakayama_2013,chen_2013} discovered in Pt\big|YIG structures a magnetoresistance, whose symmetry differs from the 
anisotropic magnetoresistance (AMR) and planar Hall effect (PHE) in magnetic thin films. This so called ``spin Hall magnetoresistance'' can be explained by the simultaneous operation of SHE and ISHE. \cite{nakayama_2013,chen_2013,hahn_2013,althammer_2013} While physically appealing, the interpretation of the observed magnetoresistance in YIG\big|bilayers in terms of the SHE is not unique. A layer of metallic Pt turned ferromagnetic by proximity would also cause different magnetoresistance effects. \cite{weiler_2012,lu_2013} However, experiments with magnetization out-of-plane show asymmetry that cannot be explained by bulk AMR, leading Lu \etal to call it a ``new magnetoresistance". \cite{lu_2013a}

In most basic terms, the experiments can be traced back to the broken inversion symmetry in these layers, irrespective of the detailed microscopic mechanism. The very simplest interface model would be a ferromagnetic 2-dimensional electron gas with broken inversion symmetry. Duine \textit{c.s.} implicitly demonstrated angular dependent transport in qualitative agreement with SMR experiments. \cite{bijl_2013} However, a ferromagnetic 2DEG is hardly a realistic model for YIG\big|Pt bilayers. Moreover, claims of the existence of a significantly magnetized Pt next to YIG have met some scepticism. \cite{geprags_2012}


In this paper we predict a magnetoresistance with the correct symmetry in normal metal\big|ferromagnetic insulator (NM\big|FI) bilayers that does not require a proximity magnetic layer nor a large spin Hall angle. The interfacial Rashba induced magnetoresistance and the spin Hall magnetoresistance have different NM film thickness dependences, therefore it is possible to distinguish these two mechanisms, at least in principle. Another way to validate our predictions would be first-principles calculations that already identified interface-enhanced spin-orbit interactions. \cite{liu_2014}



We study the charge and spin transport in a NM\big|FI bilayer structure with NM film has thickness $d$ and occupying $z\in[-d,0]$, an NM\big|FI interface at $z = 0$, and an FI with magnetization $\bM$ for $z>0$ whose thickness is irrelevant when exceeding a few monolayers. 
 The Hamiltonian of our system can be written as
\begin{equation}
H =H_{k} + H_{\bM} + U_C + H_{R}, \label{eq1}
\end{equation}
where $H_{k} =\hbp^2/2m$ is the kinetic energy with the momentum operator $\hbp$ and electron mass $m$, $H_{\bM} =J\hbsigma\cdot\bM\Theta(z)$ is the exchange interaction of the magnetization in FI and spins in NM (for weak SOI \cite{chesi_2007}) with $J$ the exchange energy and $\hbsigma$ the Pauli matrix operator, $U_C= U_+\Theta(-z-d)+U_-\Theta(z)$ is the confining potential of the NM film between vacuum and FI, $\Theta$ is the Heaviside unit step function. The last term, $H_{R} = (\eta/\hbar) \hbsigma\cdot(\hbp\times\nabla U) = \pm(\eta U/\hbar)\hbsigma\cdot(\hzz\times\hbp)\delta(z)$ is the IR-SOI due to the confining potential $U_C$ with $\eta$ the spin-orbit coupling parameter at the NM\big|FI interface, \cite{wang_2013} where the sign depends on the interface choice.


\begin{figure}[t]
\includegraphics[width=0.7\columnwidth]{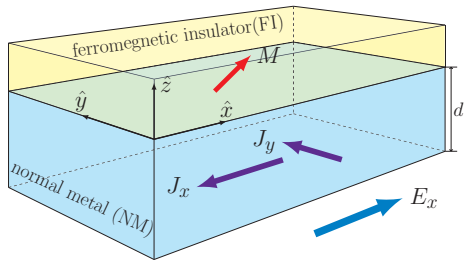} 
\caption{(Color online) Schematic picture of the system.}
\label{f1}
\end{figure}

We treat $H_{R}$ as perturbation to $H_0=H_{k}+H_{\bM}+U_C$ and for simplicity take the limit $U_+\ra\infty$ and assume $U_- \gg E_F, J$ with $E_F$ the Fermi energy. $H_0$ can be approximated as a modified quantum well in the $z$ direction with width $d$ when also $U\ra\infty$. Otherwise electrons can penetrate the FI by a depth $t = \sqrt{\hbar^2/2mU}$, hence the effective thickness of the well is $d' = d + t$. By the exchange coupling $J$ the penetration depth becomes spin dependent: $t_s = t(1 - sJ/2U)$ for spin-$s$ with $s = \pm$ and the effective thickness becomes $d_s = d+t_s$. The eigenenergies and eigenstates of $H_0$ are
\begin{align}
E_0^{ns}
&= E_{nq} + sE_n\varepsilon_J
\qwith \varepsilon_J = {tJ\ov dU}, \label{eqn:E0ns}\\
\psi_0^{ns} & = {e^{i\bq\cdot\brho}\ov\sqrt{d_s}} \sin\midb{k_{ns}(z+t_s)}\ket{s}_{\bM}, \nonumber
\end{align}
\begin{figure*}[t]
\begin{tabular}{ccc}
\hspace{0.7cm}
\includegraphics[width=.6 \columnwidth]{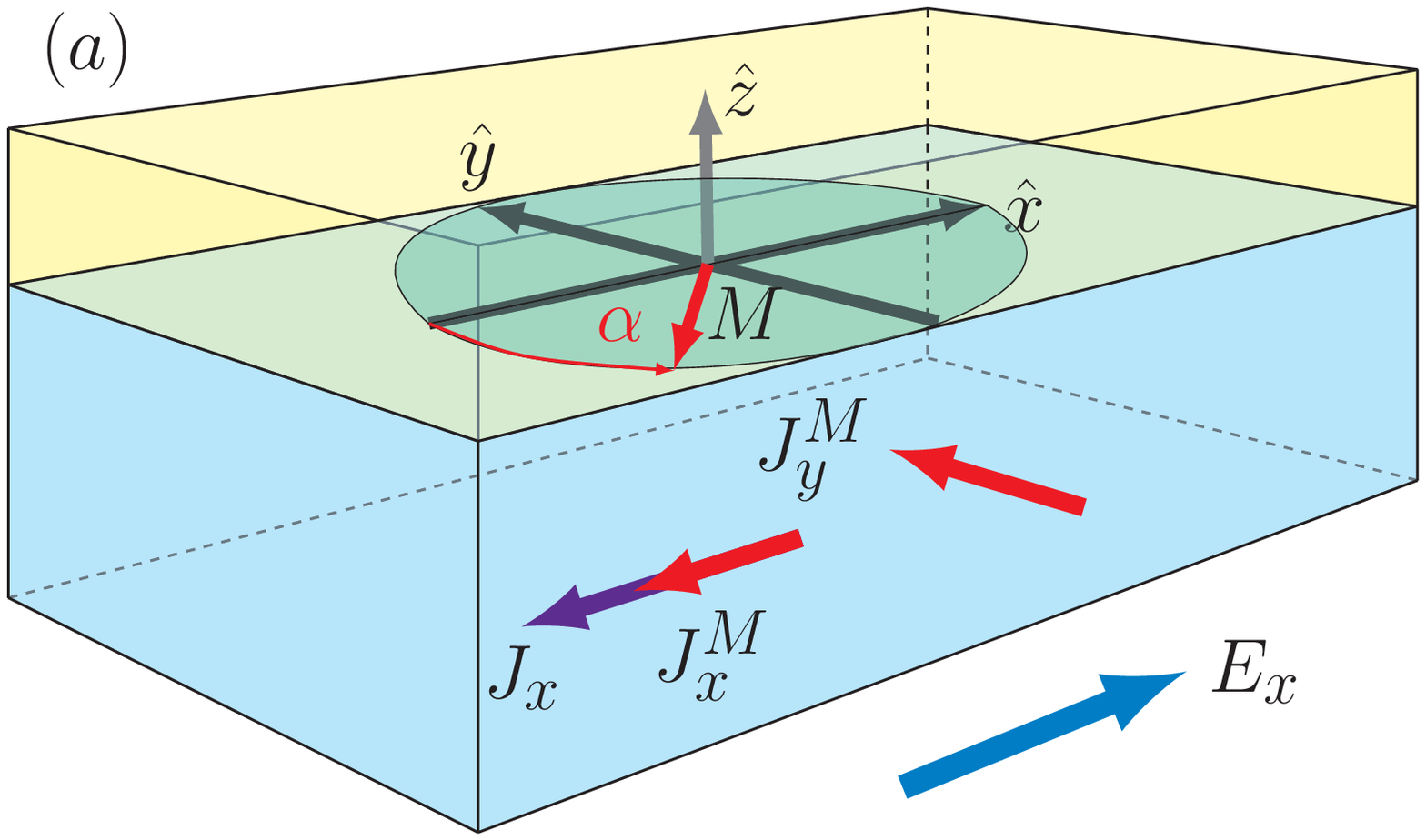}\hspace{0.5cm}
\includegraphics[width=.6 \columnwidth]{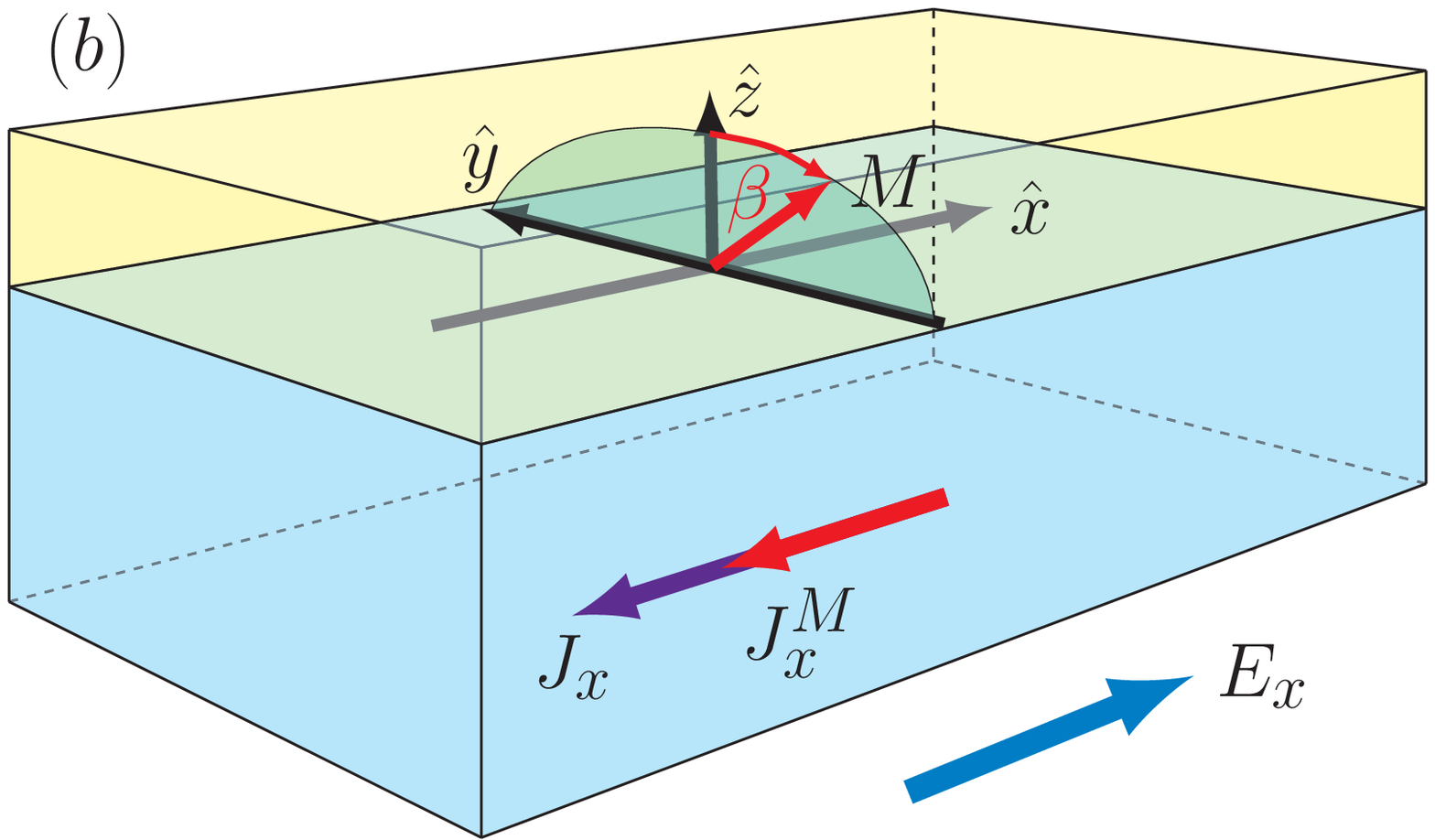} \hspace{0.5cm}
\includegraphics[width=.6 \columnwidth]{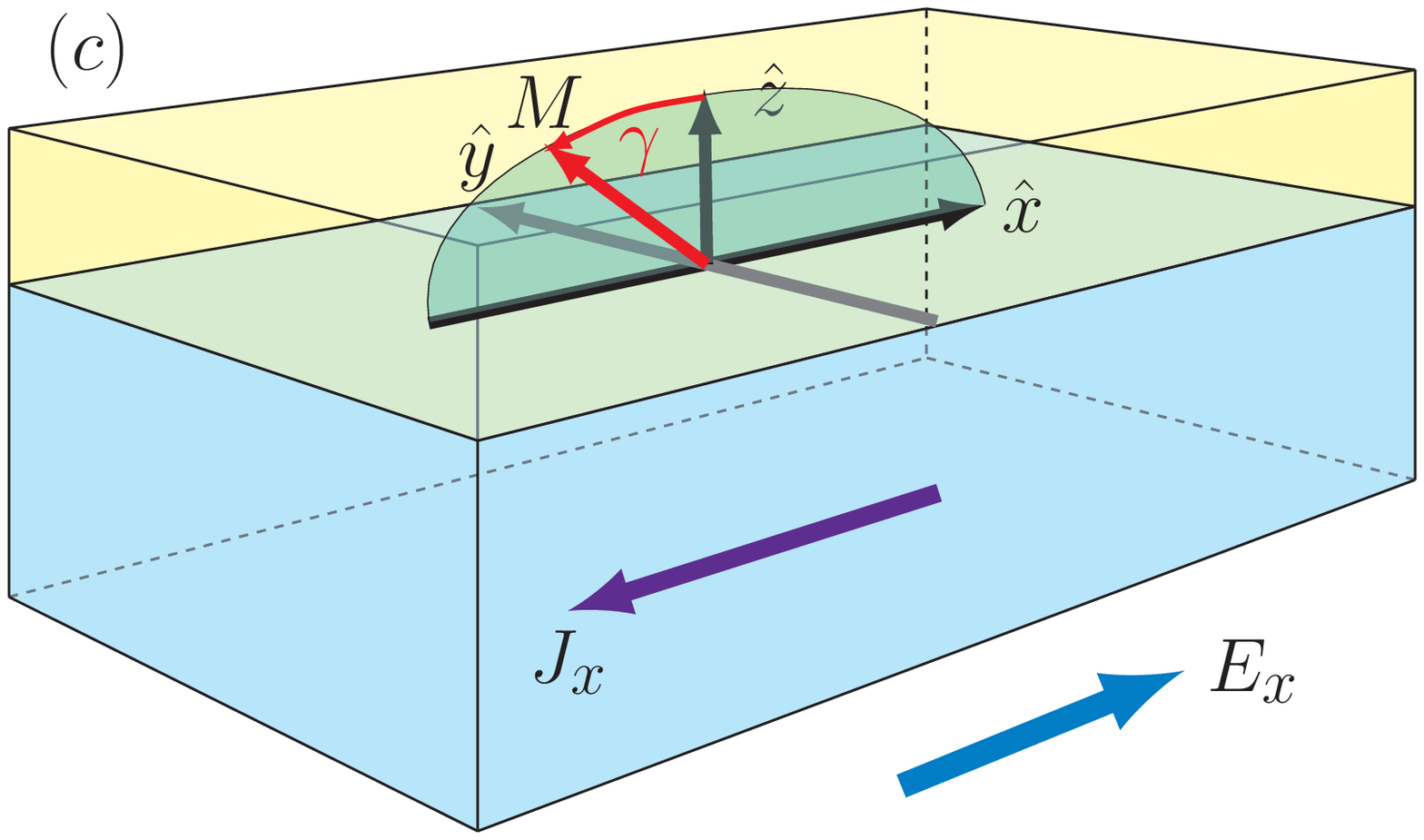}\vspace{0.5cm}\\ 
\includegraphics[width=0.96\textwidth]{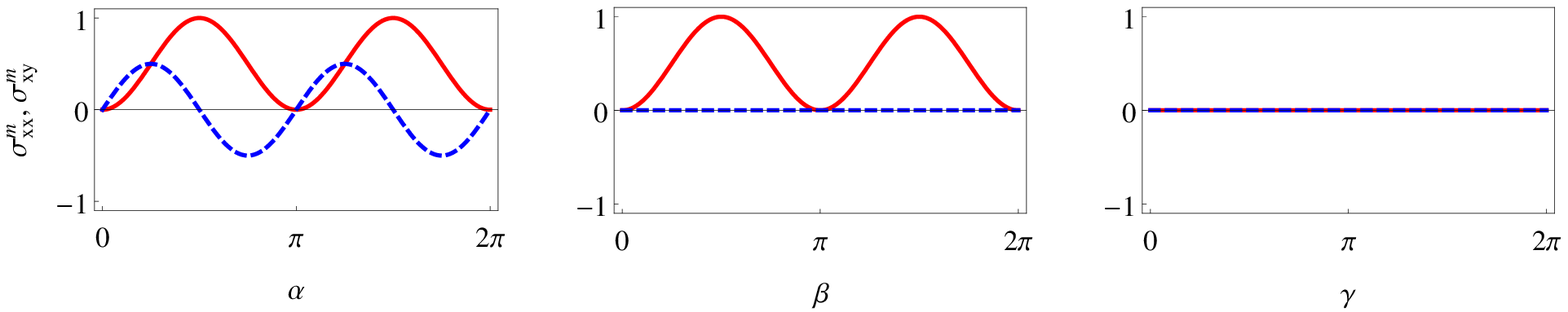}
\end{tabular} 
\caption{(Color online) Top: A schematic picture of the rotating magnetizations of FI. Bottom: the longitudinal (solid red) and transverse (dashed blue) conductivity: $\sigma_{xx}^m$ normalized by $\sigma^m_{xx} (\alpha=\pi/2)$ and $\sigma_{xy}^m$ by $\sigma^m_{xy} (\alpha=-\pi/4)$) in NM\big|FI system as a function of the magnetization angle. }
\label{fig:sigmam}
\end{figure*}
where $\brho$ and $\bq$ are the in-plane positions and wave vector, $k_{ns} = n\pi/d_s$ and $k_n = n\pi/d'\approx n\pi/d$ are the spin-dependent and the average standing wave vectors of $n$-th transverse mode, $E_n = \hbar^2k_n^2/2m$ and $E_{nq} = E_n + \hbar^2q^2/2m$, and $\ket{s}_{\bM}$ is the spinor solving $\hbsigma\cdot\bM\ket{s}_{\bM} = s \ket{s}_{\bM}$.

To linear order in $\eta$ the Hamiltonian \Eq{eq1} is diagonalized by
\begin{align}
E^{ns} 
&= E_{nq} + sE_n \smlb{\varepsilon_J - \varepsilon_{\eta}\bM\cdot{\bq_\perp\ov k_F}} + O(\varepsilon_{\eta}^2), \label{eqn:Ens} \\
\psi^{ns} &= \psi_0^{ns} - {s\varepsilon_{\eta} \ov 2 \varepsilon_J}
\hbsigma_{\bar{s}s}\cdot{\bq_\perp\ov k_F}\psi_0^{n\bar{s}}
\qwith \varepsilon_{\eta} = {d_{\eta}\ov d} = {\eta k_F\ov d}, \label{eqn:psins} \nonumber
\end{align}
where $\bq_\perp = \hzz\times\bq$, $\bar{s}=-s$, and $\hbsigma_{s's}= \brak{s'}{\hbsigma}{s}_\bM$. The degeneracy in \Eq{eq1} is now lifted because of i) the exchange coupling with the magnetization in FI: the $\varepsilon_J$ term in \Eq{eqn:Ens} and ii) the spin-orbit interaction at the surface, the $\varepsilon_{\eta}$ term in \Eq{eqn:Ens}. The latter does not depend on the barrier height $U$ because a larger potential gradient for spin-orbit interaction and smaller probability of finding the electron at the interface cancel each other, which implies that the effects predicted here do not depend sensitively on the details of the model. The interfacial Rashba spinor dominantly mixes opposite spins in subbands with the same quantum numbers $n$ and $\bq$. 

The in-plane velocity operator $\hbv_\|$ 
\begin{equation}
\hbv_\| = -{i\ov\hbar}[\brho,H]
= {\hbp\ov m} + { \eta U \ov \hbar}\delta(z)\hbsigma\times\hzz, 
\label{eqn:vpar}
\end{equation}
acquires an anomalous component. Therefore the expectation value average of the velocity over the state $\psi^{ns}$ is 
\begin{align}
\label{eqn:jns}
&\bv^{ns}(\bq) = \brak{\psi^{ns}}{\hbv_\|}{\psi^{ns}} = {1\ov\hbar}{\partial E^{ns}\ov\partial\bq}
= {\hbar\bq\ov m} \\
&+ {sE_n\varepsilon_{\eta} \ov \hbar k_F}
\bigb{\bM\times\hzz + {\varepsilon_{\eta}\ov\varepsilon_J}
\rem{\smlb{\hbsigma_{s\bar{s}}\cdot{\bq_\perp\ov k_F} } (\hbsigma_{\bar{s}s} \times \hzz) } }, \nonumber
\end{align}
where the first term is the normal velocity, the second term is the IR-SOI induced anomalous velocity up to second order in $\varepsilon_{\eta}$. 
The spin-$\pm$ is boosted in the $\pm\bM\times\hzz$ direction due to the IR-SOI induced anomalous velocity. Considering that the population for spin-$\pm$ differ due to the energy splitting in \Eq{eqn:Ens}, an additional charge current flows in the $\bM\times\hzz$ direction.


With the qualitative understanding above, we now calculate the charge current driven by an in-plane electric field $\bE$. The electric field introduces a shift in the Fermi surface $\delta\bq = e\bE\tau/\hbar$, where $\tau$ is the electron relaxation time. We assume the bulk impurities dominate the relaxation, such that $\tau$ is a constant for all subbands. \cite{zhou,jin_2007} The  electric field-induced drift is reflected by the distribution function
\begin{equation}
\label{eqn:gns}
g_{ns}(\bq)
= {e\tau\ov \hbar}\delta\smlb{E^{ns}-E_F}{\partial E^{ns}\ov\partial \bq}\cdot\bE.
\end{equation}
The total charge current density $\bJ = e\sum_{ns} \int d^2\bq~g_{ns}(\bq)\bv^{ns}(\bq)$
is calculated by expanding \Eqs{eqn:jns}{eqn:gns} to second order in $\varepsilon_{\eta},$ noting that odd terms in $s$ and $\bq$ vanish. The charge current $\bJ = \bJ_{\parallel} + \bJ_{\perp}$ contains longitudinal $\bJ_{\parallel}$ and (in-plane) transverse components $\bJ_\perp$ to the applied electric field. $\bE = E_x \hxx$, then $\bJ_\| = J_x \hxx$ and $\bJ_\perp = J_y \hyy$:
\begin{subequations}
\label{eqn:Jxy}
\begin{align}
J_x &= \sigma_{xx} E_x = 
\midb{ {d\ov d_{\eta}} + {3\ov 20}{d_{\eta}\ov d}\smlb{2 + M_y^2}} d_{\eta} \sigma_0 E_x, \\
J_y &= \sigma_{xy} E_x = \smlb{-{3\ov 20}{d_{\eta}\ov d} M_xM_y} d_{\eta} \sigma_0 E_x, 
\end{align}
\end{subequations}
with $\sigma_0 = (k_F^3/3\pi^2)(e^2\tau/m)$ being the Drude conductivity of bulk NM. \Eq{eqn:Jxy} is our main result.

\Eq{eqn:Jxy} depends on the magnetization direction, thereby predicting an SOI induced magnetoresistance (SMR). When the magnetization rotates in the $\hxx$-$\hyy$ plane, both the longitudinal and transverse components of the SMR oscillates with an amplitude of $\sigma_{xx}^m = \sigma_{xy}^m = (3\varepsilon_\eta/20) (d_\eta\sigma_0)$. In \Figure{fig:sigmam}, we plot $\sigma_{xx}^m$ and $\sigma_{xy}^m$ as a function of angle $\alpha,$ $\beta$ and $\gamma$ of the magnetization in FI layer in $x$-$y$, $y$-$z$ and $x$-$z$ planes, respectively. The longitudinal magnetoresistance depends on angle $\alpha$ and $\beta$, but not on $\gamma$, while the transverse magnetoresistance depends only on $\alpha,$ in contrast to the bulk metal anisotropic magnetoresistance (AMR), \cite{mcguire_1975} which depends only on the relative angle between the magnetization and the charge current, thus $M_x,$ \textit{i.e.} on $\beta$, but not on $\gamma$. However, the angle dependence of SMR shown in \Figure{fig:sigmam} agrees with the experiments. \cite{nakayama_2013,chen_2013,hahn_2013,althammer_2013} The failure of our model to find an anomalous Hall-like effect is not an important worry because it is observed \cite{vlietstra_2013} to be magnitudes smaller than the planar Hall-like effect. 

\begin{figure}[t]
\includegraphics[width=0.7\columnwidth]{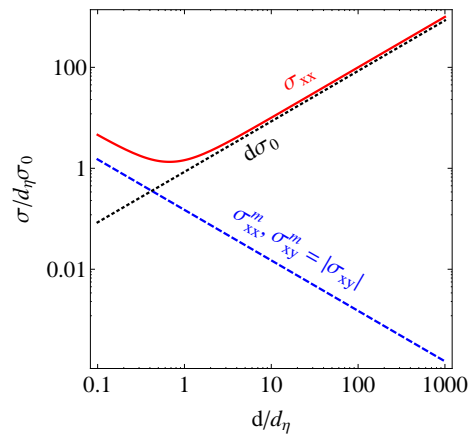} 
\caption{(Color online) The NM film thickness $d$ dependence of the longitudinal conductivity $\sigma_{xx}$ for $\bM\|\hyy$ ($\alpha=-\pi/2$, solid curve), the longitudinal/transverse magneto-conductivity $\sigma_{xx}^m = \sigma_{xy}^m = \abs{\sigma_{xy}}$ (dashed curve) and the limiting longitudinal conductivity for large $d$ (dotted curve). The longitudinal conductivity has a minimum at $d/d_{\eta} = \sqrt{9/20}$.
} \label{fig:sigmad}
\end{figure}

Since the angular dependence is identical, the origin of the recently experimentally discovered ``SMR'' is still debatable: it can be caused by the bulk spin Hall effects as well as by an interfacial Rashba effect. In principle, different dependences on the NM film thickness may be expected. In \Figure{fig:sigmad}, we plot our results for the longitudinal and transverse conductivity as function of the scaled NM film thickness $d/d_\eta$. The longitudinal conductivity (solid red curve in \Figure{fig:sigmad}) has a minimum at $d/d_\eta = \sqrt{9/20}$ because the bulk contribution is proportional to $d$, while the IR-SOI contribution is proportional to $1/d$. However, the conductvity minimum can only be observed for interfaces with very strong IR-SOI (large $\eta$) because $d_\eta = \eta k_F$ is smaller than the lattice constant for typical values of $\eta k_F^2 \sim 0.1$. On the other hand, both the planar Hall conductivity and the longitudinal magneto-conductivity are proportional to $1/d$, thus increase with decreasing thickness.

From \Eq{eqn:Jxy}, the ratio between the magnetoconductance and the average conductance:    
\begin{equation}
{\sigma_{xx}^m\ov \bar{\sigma}_{xx}} = {3\ov 20}\varepsilon_{\eta}^2 \qRa 
\eta  = {d\ov k_F}\sqrt{{20\ov 3}{\sigma_{xx}^m\ov \bar{\sigma}_{xx}}},
\end{equation}
therefore it is possible to estimate the interfacial Rashba spin-orbit interaction parameter $\eta$ from transport measurements. These estimates are valid for films that are thinner than the spin-flip diffusion length. In the same regime the magnetoresistance caused by the spin $~$Hall effects on the metal vanishes identically \cite{chen_2013}, thereby predicts qualitatively different behavior. As a note of caution we point out that surface roughness and interface spin-flip scattering, which are not taken into account in either theory, might complicate the interpretation of experiments.




In conclusion, we predict a magnetoresistance in an NM\big|FI bilayer system induced by the interfacial Rashba spin-orbit interaction. This new magnetoresistance has the same symmetry on the magnetization direction as the spin Hall magnetoresistance. We therefore question the physical origin of the recently discovered spin Hall magnetoresistance. A NM film thickness dependence measurement can clarify which effect, the interfacial Rashba or the bulk spin Hall effect, is responsible. We propose that rather than ``spin Hall magnetoresistance" or ``new magnetoresistance" the effect should be called ``spin-orbit magnetoresistance". 

This work was supported by the special funds for the Major State Basic Research Project of China (2014CB921600, 2011CB925601), the National Natural Science Foundation of China (91121002), the Foundation for Fundamental Research on Matter (FOM), Marie Curie ITN Spinicur, DFG Priority Programme 1538 "Spin-Caloric Transport", Grant-in-Aid for Scientific Research (Kakenhi) 25247056/25220910, and EU-FET Grant InSpin 612759.


\end{document}